\documentclass[aps,pra,showpacs,superscriptaddress,reprint]{revtex4-1}

\usepackage{amsmath,amssymb}
\usepackage{natbib}
\usepackage{graphicx}
\usepackage{bm}
\usepackage{braket}
\usepackage{color}

\newcommand{\halfphi}{\left(\frac{\phi}{2}\right)}
\newcommand{\rms}{\mathrm{s}}
\newcommand{\rma}{\mathrm{a}}
\newcommand{\nbar}{\bar{n}}
\newcommand{\hopratio}{\left(\frac{J_{\perp}}{2J}\right)}

\def \plotwi{8.3cm}

\begin{document}
\title{Ground States of a Bose-Hubbard Ladder 
in an Artificial Magnetic Field:\\Field-Theoretical Approach}

\author{Akiyuki Tokuno}
\affiliation{Centre de Physique Th\'eorique, Ecole Polytechnique, CNRS, 91128 Palaiseau Cedex, France}
\affiliation{Coll\`{e}ge de France, 11 place Marcelin Berthelot, 75005 Paris, France.}
\author{Antoine Georges}
\affiliation{Centre de Physique Th\'eorique, Ecole Polytechnique, CNRS, 91128 Palaiseau Cedex, France}
\affiliation{Coll\`{e}ge de France, 11 place Marcelin Berthelot, 75005 Paris, France.}
\affiliation{DPMC, University of Geneva, 24 quai Ernest-Ansermet, CH-1211 Geneva, Switzerland.}

\begin{abstract} 
 We consider a Bose-Hubbard ladder subject to an artificial magnetic
 flux and discuss its different ground states, their physical
 properties, and the quantum phase transitions between them.
 A low-energy effective field theory is derived, in the two distinct
 regimes of a small and large magnetic flux, using a bosonization
 technique starting from the weak-coupling limit.
 Based on this effective field theory, the ground-state phase diagram at
 a filling of one particle per site is investigated for a small flux and
 for a flux equal to $\pi$ per plaquette.
 For $\pi$-flux, this analysis reveals a tricritical point which has
 been  overlooked in previous studies.
 In addition, the Mott insulating state at a small magnetic flux is
 found  to display Meissner currents.
\end{abstract}


\maketitle


\section{Introduction}
Recent developments in ultra-cold atom physics allow for studies of a
wide range of many-body quantum systems of bosons, fermions and their
mixtures, involving  strong correlations and/or frustration. 
One of the remarkable recent advances is the so-called artificial gauge
fields,~\cite{Dalibard.et.al/RMP83.2011} which allow one to generate
spin-orbit couplings and magnetic fields, opening the way for example to
quantum Hall and spin Hall effects.
These effects are also related to studies on topological phases of
matters.
In addition, the control of interactions between atoms by the Feshbach
resonance technique allows for the controlled study of quantum systems
under the combined effects of an artificial gauge field and strong
correlations.

The key to artificial gauge fields is 
Berry's phases~\cite{Berry/PRS392.1984} tuned by atom-light 
interactions,~\cite{Dum.Olshanii/PRL76.1996} in which atoms acquire a 
geometric phase in their motion because of an adiabatic spatial
change of the dressed states.
Using Raman transitions based on these ideas, the synthesis of an 
effective magnetic field~\cite{Lin.et.al/Nature462.2009} and spin-orbit
coupling~\cite{Lin.et.al/Nature471.2011,Wang.et.al/PRL109.2012,Cheuk.et.al/PRL109.2012}
have been experimentally achieved in Bose condensates of $^{87}$Rb
atoms, and the spin-Hall effect in Bose condensates has been also
successfully observed~\cite{Beeler.et.al/Nature498.2013}. 

The realization of artificial gauge fields in optical lattice potentials
has also been intensively discussed.
The pioneering theory making use of photo-assisted tunneling techniques
has been established by Jaksch and
Zoller~\cite{Jaksch.Zoller/NJP5.2003}.
Subsequently other schemes for effective uniform magnetic
fields~\cite{Sorensen.et.al/PRL94.2005,Mueller/PRA70.2004} and for
staggered magnetic
fields~\cite{Lim.Smith.Hemmerich/PRL100.2008,Lim.et.al/PRA81.2010} have also been proposed. 
In experiments, several types of artificial magnetic fields using
photo-assisted tunneling have been subsequently realized in recent
years: effective magnetic fluxes inhomogeneously set in
stripes~\cite{Aidelsburger.etal/PRL107.2011,Aidelsburger.etal/AppPhysB113.2013}, 
uniform magnetic fluxes~\cite{Miyake.etal/PRL111.2013}, and  spin-orbit
couplings without spin 
flips~\cite{Kennedy.et.al/PRL111.2013,Aidelsburger.etal/PRL111.2013} in
two-dimensional optical lattice systems.
In addition to the above realizations, other schemes for artificial
gauge fields have been invented using Zeeman lattice 
techniques~\cite{JimenezGarcia.et.al/PRL108.2012} and shaking of optical
lattice
potentials~\cite{Struck.etal/PRL108.2012,Simonet.et.al/EPJ57.2013,Struck.et.al/NatPhys9.2013}.

Optical lattices also allow for the control of dimensionality, so that 
one-dimensional quantum systems can be realized. 
A quasi-one-dimensional ``ladder'' geometry plays the role of a
minimal model for studying the effect of gauge fields. 
In these low-dimensional systems, the whole range of interaction strengths
from weak to strong coupling can be investigated using powerful
numerical and analytic techniques, such as bosonization and the
density-matrix renormalization group (DMRG).
Because of the peculiar critical nature of Tomonaga-Luttinger (TL)
liquids which describe their low-energy properties, such
quasi-one-dimensional quantum systems subject to artificial gauge fields
or high magnetic fields are expected to display interesting
phenomena.
In studies on ladder systems subjected to magnetic fields, fermion
systems have been discussed in the context of strongly correlated electron
systems~\cite{Carr.Narozhny.Nersesyan/PRB73.2006,Jaefari.Fradkin/PRB85.2012,
Carr.Narozhny.Nersesyan/PRB73.2006,Roux.Orignac.White.Poiblanc/PRB76.2007,
Schollwock.et.al/PRL90.2003}. 
The study of bosonic ladders subject to magnetic fields has also been
motivated by the Josephson junction ladders and ultra-cold Bose atoms in
optical
lattices~\cite{Denniston.Tang/PRL75.1995,Kardar/PRB33.1986,Granato/PRB42.1990,Nishiyama/EPJB17.2000,Orignac.Giamarchi/PRB64.2001,Petrescu.LeHur/PRL111.2013,Dhar.etal/PRB87.2013,Dhar.etal/PRA85.2012,Crepin.et.al/PRB84.2011,Tovmasyan.et.al/PRB88.2013}. 
In addition to common features of Bose-Hubbard models such as a
one-dimensional superfluid (SF), Mott insulator (MI), and phase
transition between them, the bosonic ladders exhibit interesting
phenomena induced by the magnetic field: chiral superfluid phases (CSF)
and chiral Mott insulating phases (CMI) displaying Meissner
currents~\cite{Orignac.Giamarchi/PRB64.2001,Dhar.etal/PRB87.2013,Dhar.etal/PRA85.2012,Petrescu.LeHur/PRL111.2013}.
Attention to the topic of bosonic ladders subject to an artificial
magnetic field has been reinforced recently by the first experimental
realization of such a system.~\cite{Atala.et.al/arXiv2014}

In this article, we study the low-energy physics of Bose-Hubbard ladders 
subject to an artificial uniform magnetic flux, from the viewpoint of
field theory.
So far, field theoretical approaches to bosonic ladder systems have
usually considered starting with the strong coupling limit, in which the
rung hopping is treated
perturbatively~\cite{Donohue.Giamarchi/PRB63.2001,Orignac.Giamarchi/PRB64.2001,Petrescu.LeHur/PRL111.2013}
In contrast, we derive an effective field theory from a 
weak coupling perspective, in which the effect of a rung hopping is 
fully taken into account, and the interaction is included perturbatively. 
In this approach, typical strong correlation phenomena such as the MI
state and the MI-SF phase transition can nonetheless be investigated, by
taking into account backscattering and umklapp scattering processes. 
The low-energy effective field theories in the two cases of a large and
small magnetic flux are separately derived, for an
arbitrary filling.
In addition we also apply the constructed effective field theory and
investigate the ground-state phase diagram in two limiting cases, namely
that of a large magnetic flux equal to $\pi$ per plaquette, and that of
a small magnetic flux, with one particle per site.
For the $\pi$ magnetic flux, we compare our phase diagram to the one
previously obtained numerically in Ref.~\cite{Dhar.etal/PRA85.2012}, and
all the phases found there are well described  by our approach.
Furthermore, more importantly, the presence of a tricritical point in
the phase diagram is predicted by the analysis presented here, which
has not been emphasized previously.
In the limit of a small magnetic flux, we show that a SF state with
Meissner current appears, and transits to the MI state for strong
interactions.
In addition we also find that the Meissner current can survive also in
the MI state while the system is fully gapped. 
A similar fully gapped charge-ordered state with Meissner currents has
also been found for the different filling of one particle per two sites
by Petrescu and Le Hur~\cite{Petrescu.LeHur/PRL111.2013}.

This paper is organized as follows.
In Sec.~\ref{sec:model_and_effective_theory} we first define the
considered Bose-Hubbard ladder with a magnetic flux.
Next we analyze the single-particle band structure in the
non-interacting case in Sec.~\ref{sec:single-particle-spectrum}, and
derive the general form of the low-energy effective field theory for
a large magnetic flux in Sec.~\ref{sec:effective_theory_large_flux},
and for a small magnetic flux in Sec.~\ref{sec:low-magnetic_flux}.
Furthermore, based on the derived field theory, we investigate the
ground-state phase diagrams in Sec.~\ref{sec:discussion}.
A summary and conclusion are provided in Sec.~\ref{sec:summary}.
In Appendix~\ref{sec:MF}, the mean-field analysis which is used to
construct the effective field theories is presented.

\section{Model and effective theory}\label{sec:model_and_effective_theory}
\begin{figure}[tbp]
 \centering
 \includegraphics[scale=1.]{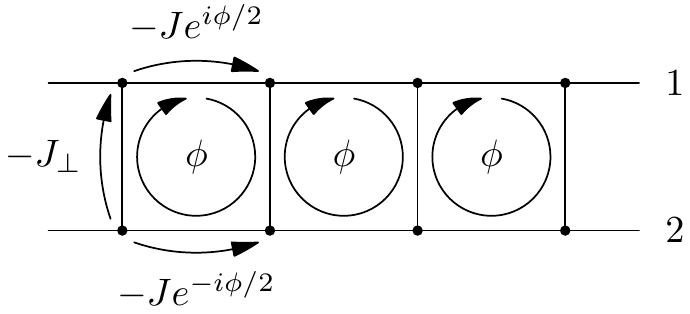}
 \caption{The Bose-Hubbard ladder Hamiltonian considered in this paper.
 Due to the magnetic field, hopping involves a phase factor associated
 with the corresponding gauge field.
 The phases gained in the hopping processes are displayed, corresponding
 to the gauge choice defined by Eq.~(\ref{eq:gauge_choice}).}
 \label{fig:ladder}
\end{figure}

Let us define the Bose-Hubbard ladder Hamiltonian with an applied
uniform artificial magnetic field,
\begin{align}
 H&=H_{0}+H_{\mathrm{loc}},
 \label{eq:H}\\
 H_{0}
 &= -J\sum_{p=1,2}\sum_{j}
      \left[
        e^{iA^{\parallel}_{j,p}}b^{\dagger}_{j+1,p}b_{j,p}
        +\mathrm{h.c.}
      \right]
 \nonumber \\
 & \quad
    -J_{\perp}\sum_{j}
      \left[
        e^{iA^{\perp}_{j}}b^{\dagger}_{j,1}b_{j,2} + \mathrm{h.c.}
      \right],
 \label{eq:H_0}\\
 H_{\mathrm{loc}}
 &=\sum_{j}\sum_{p=1,2}
   \left[
     -\mu n_{j,p}
     +\frac{U}{2}n_{j,p}\left(n_{j,p} + 1\right)
   \right],
 \label{eq:H_U}
\end{align}
where $n_{j,p}=b^{\dagger}_{j,p}b_{j,p}$ is a number operator, and the
index $p=1,2$ denotes the upper and lower chain, respectively.
The model Hamiltonian is schematically illustrated in
Fig.~\ref{fig:ladder}.
This model has the two different hoppings, intrachain $J>0$ and
interchain $J_{\perp}>0$, and only a repulsive on-site Hubbard
interaction $U>0$ is considered.
The artificial magnetic field is introduced via the Peierls
substitution, and the corresponding gauge field along the chain
direction on the chain $p$, and along the rung direction are denoted by
$A^{\parallel}_{j,p}$ and $A^{\perp}_{j}$, respectively. 
This produces the applied artificial magnetic flux $\phi$ piercing a
plaquette as 
\begin{equation}
 \oint_{\Box}\bm{A}\cdot d\bm{l}
 = A^{\parallel}_{j,1}-A^{\perp}_{j+1}-A^{\parallel}_{j,2}+A^{\perp}_{j}
 = \phi.
 \label{eq:flux_condition}
\end{equation}
In this paper, we choose the following gauge:
\begin{equation}
 \left\{
 \begin{aligned}
  & A^{\parallel}_{j,1}=\phi/2, \\
  & A^{\parallel}_{j,2}=-\phi/2, \\
  & A^{\perp}_{j}=0,
 \end{aligned}
 \right.
 \label{eq:gauge_choice}
\end{equation}
which obviously obeys Eq.~(\ref{eq:flux_condition}).
The Hamiltonian (\ref{eq:H}) is invariant under the transformation,
$(\phi,b_{j,1},b_{j,2})\rightarrow(-\phi,b_{j,2},b_{j,1})$.
Thus the magnetic flux $\phi$ to be considered can be primitively
reduced, and we restrict the magnetic flux to be $0<\phi\le\pi$
throughout this paper.

\subsection{Single-particle energy band structure}
\label{sec:single-particle-spectrum}
\begin{figure*}
 \centering
 \includegraphics[scale=1.]{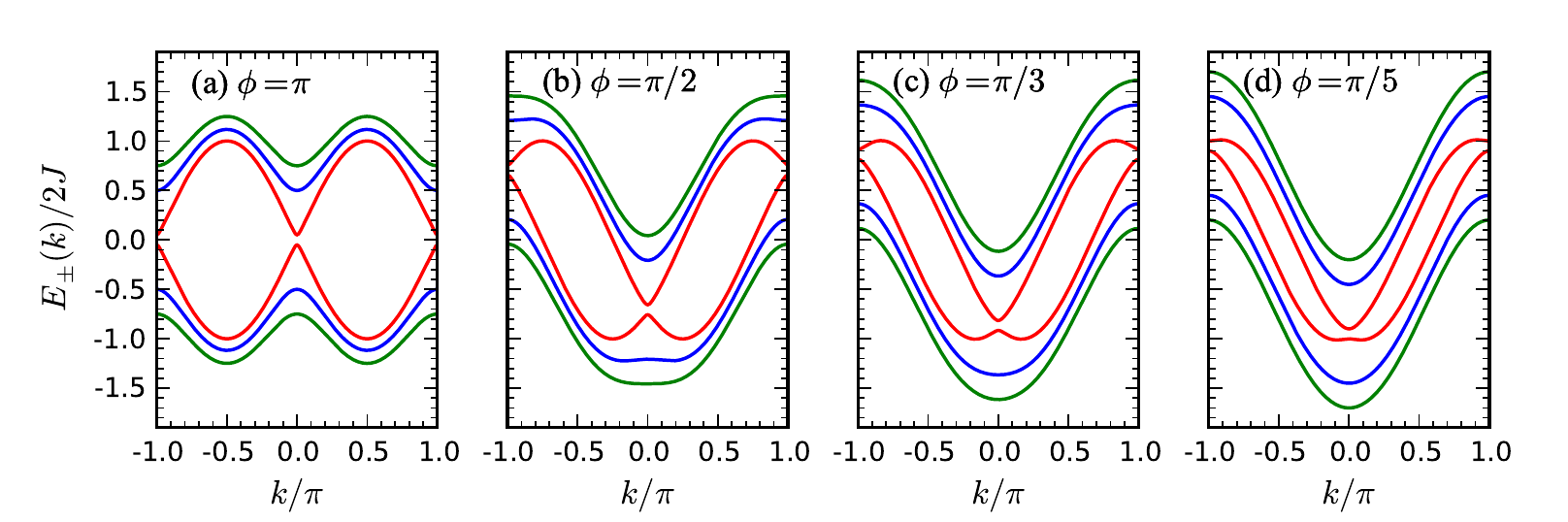}
 \caption{(Color online) The single-particle energy bands of the
 two-leg ladder with a uniform magnetic flux: (a) $\phi=\pi$, (b)
 $\phi=\pi/2$, (c) $\phi=\pi/3$ and (d) $\phi=\pi/5$. 
 The red, blue and green lines denote $J_{\perp}/2J=0.05$, $0.5$ and
 $0.75$, respectively.
 Here only the commensurate magnetic fluxes are shown, but the structure
 of the spectrum is continuously deformed by varying the flux $\phi$.}
 \label{fig:bands}
\end{figure*}

Let us look at the single-particle spectrum. 
The single-particle Hamiltonian~(\ref{eq:H_0}) is easily diagonalized by
the following unitary transformation as
\begin{equation}
 \left\{
  \begin{aligned}
   & b_{1}(k)=v_{k}\alpha(k)+u_{k}\beta(k),\\
   & b_{2}(k)=-u_{k}\alpha(k)+v_{k}\beta(k), \\
  \end{aligned}
 \right.
 \label{eq:u-trans}
\end{equation}
where $b_{p}(k)$ with $p=1,2$ is a Fourier transformation of $b_{j,p}$.
The coefficients are given as
\begin{equation}
 \left\{
 \begin{aligned}
  u_{k}
  &=\sqrt{\frac{1}{2}\left(1-\frac{\sin(\phi/2)\sin{k}}{\sqrt{(J_\perp/2J)^2+\sin^{2}(\phi/2)\sin^{2}{k}}}\right)},
  \\
  v_{k}
  &=\sqrt{\frac{1}{2}\left(1+\frac{\sin(\phi/2)\sin{k}}{\sqrt{(J_\perp/2J)^2+\sin^{2}(\phi/2)\sin^{2}{k}}}\right)}.
 \end{aligned}
 \right.
\end{equation}
Consequently the single-particle Hamiltonian~(\ref{eq:H_0}) has a 
two-band structure:
\begin{align}
 H_0
 &= \sum_{k}
    \left[
     E_{+}(k)\alpha^{\dagger}(k)\alpha(k)
     + E_{-}(k)\beta^{\dagger}(k)\beta(k)
    \right],
\end{align}
with the single-particle energy bands being
\begin{align}
 E_{\pm}(k)
 &=-2J \cos\left(\frac{\phi}{2}\right)\cos{k}
 \nonumber \\
 &\quad
 \pm
 \sqrt{J_{\perp}^2+(2J)^2\sin^2\left(\frac{\phi}{2}\right)\sin^2{k}}.
 \label{eq:enegy_bands}
\end{align}
The energy dispersions $E_{\pm}(k)$ correspond to the upper and lower
band, respectively. 
The band structures for certain values of $\phi$ and $J_{\perp}/J$ are
shown in Fig.~\ref{fig:bands}.
Comprehensive results on the dependence of the band structure on the
magnetic field $\phi$ and hopping ratio $J_{\perp}/J$ can be found in
Refs.~\cite{Carr.Narozhny.Nersesyan/PRB73.2006,Roux.Orignac.White.Poiblanc/PRB76.2007}.  
In addition, the band structure of the Hamiltonian $H_{0}$ is
known to be analogous to that of spin-$1/2$ particles with a spin-orbit
coupling in the presence of a magnetic field~\cite{Huegel.Paredes/PRA89.2014}.

\begin{figure}
 \centering
 \includegraphics[width=\plotwi]{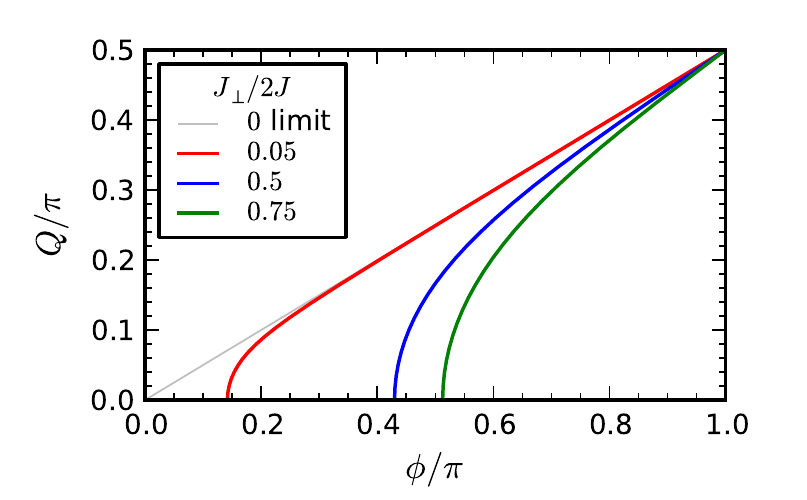}
 \caption{(color online)
 The position of the momentum corresponding to the band minima,
 $k_{\min}=\pm Q$, as a function of the magnetic flux per plaquette
 $\phi$ for several hopping ratios $J_{\perp}/2J$. 
 A finite $Q$ corresponds to the case when the lower band displays two
 minima, while $Q=0$ means that the band has a single minimum
 structure.
 The critical value of the magnetic flux, at which $Q$ becomes zero,
 increases with $J_\perp/J$ as in Eq.~(\ref{eq:critical_flux}).}
 \label{fig:band_minima}
\end{figure}

The band structure around the lowest energy is the most important
feature for low-energy physics, since bosons tend to populate states
around energy minima. 
Thus we focus here only on the features at the bottom of the lower band.
In the regime of a large magnetic flux per plaquette the lower band
$E_{-}(k)$ has two separate minima, and the corresponding wave numbers
$k_{\mathrm{min}}$ at the band minima are given analytically as
$k_{\mathrm{min}}=\pm Q$ with 
\begin{equation}
 Q=\arccos
   \left[
    \cot\left(\frac{\phi}{2}\right)
    \sqrt{\hopratio^2+\sin^2\left(\frac{\phi}{2}\right)}
   \right],
\end{equation}
which is shown in Fig.~\ref{fig:band_minima}.
The two band minima are maximally separated for $\phi=\pi$, and located
exactly at $k_{\mathrm{min}}=\pm\pi/2$.
As the flux $\phi$ decreases, the two minima approach one another, and
eventually merge at a critical value of the magnetic flux
\begin{equation}
 \phi_{\mathrm{c}}
  =2\arccos\left[\sqrt{\left(\frac{J_\perp}{4J}\right)^2+1}-\frac{J_\perp}{4J}\right].
  \label{eq:critical_flux}
\end{equation}
A single minimum structure is formed for small $\phi$.
On the other hand, the hopping ratio $J_{\perp}/J$ works so as to
enlarge the distance between the two bands, and so as to narrow the band
width.
Thus the increase of the ratio $J_{\perp}/J$ suppresses the height of the 
barrier between the two minima in the lower band, which also leads to the
increase of the critical $\phi_{c}$ with $J_{\perp}/J$.
In what follows, we separately consider the two different cases.
The first is the case of a sufficiently large magnetic flux
$\phi\gg\phi_\mathrm{c}$, in which the bottom of the lower band exhibits
double minima, and they are clearly separated. 
The second is the case of a small  magnetic flux $\phi<\phi_c$, in which
the band bottom forms a single minimum structure.

\subsection{Effective Hamiltonian for large magnetic flux}
\label{sec:effective_theory_large_flux}
Let us derive the low-energy effective theory based on the
single-particle spectrum obtained above. 
For a large enough magnetic flux $\phi\gg\phi_\mathrm{c}$, we have a 
double-minimum structure in the lower energy band, and in the ground
state the bosons dominantly populate the two energy minima even in the
presence of the interaction.
Thus, the spectrum relevant to the low-energy physics can be 
approximated by expanding the band-structure around the energy minima:
\begin{align}
 E_{-}(k=\pm Q+q)\approx -E_{0}+\frac{q^2}{2m^{*}},
\end{align}
where $E_{0}=-E(k=\pm Q)$ and $m^{*}$ are, respectively, a minimum
energy offset and effective mass, and they are given as
\begin{align}
 & E_{0}
 = \frac{2J}{\sin\halfphi}\sqrt{\hopratio^2+\sin^2\halfphi},
 \\
 & \frac{1}{m^{*}}=\frac{d^2E(k=\pm Q)}{dk^2}.
\end{align}
The wave number $q$ denotes the variation from the minima
$k_{\mathrm{min}}=\pm Q$, and is assumed to be small enough, $q\ll 1$.
The minimum energy offset $-E_{0}$ shifts the chemical potential $\mu$
in the Hamiltonian~(\ref{eq:H}).
Thus under this long-wave-length approximation we need to fix the
chemical potential including this energy offset to reproduce the
required density.

Correspondingly to the long-wave-length expansion, the unitary
transformation~(\ref{eq:u-trans}) is also approximated as follows. 
For the upper chain, 
\begin{equation}
 \left\{
 \begin{aligned}
  & b_{1}(k=Q+q) 
  = V_{-}\beta_{+}(q),
  \nonumber \\
  & b_{1}(k=-Q+q) 
  = V_{+}\beta_{-}(q),
 \end{aligned}
 \right.,
\end{equation}
and for the lower chain,
\begin{equation}
 \left\{
 \begin{aligned}
  & b_{2}(k=Q+q)
  = V_{+}\beta_{+}(q),
  \nonumber \\
  & b_{2}(k=-Q+q)
  = V_{-}\beta_{-}(q).
 \end{aligned}
 \right.,
\end{equation}
where the weight factors, $V_{+}=u_{-Q}=v_{Q}$ and $V_{-}=u_{Q}=v_{-Q}$,
have been introduced, and they are explicitly given as
\begin{equation}
 V_{\pm}
 = \sqrt{\frac{1}{2}
    \left[
     1\pm\sqrt{\frac{\sin^2(\phi/2)-(J_\perp/2J)^2\cot^2(\phi/2)}{(J_\perp/2J)^2+\sin^2(\phi/2)}}
    \right]
   }.
\end{equation}
This approximation means that all the energy states except for
the low-energy states near the band bottom are projected out.

The approximated boson operators are represented in real space as
\begin{equation}
 \left\{
 \begin{aligned}
 & b_{j,1}
 \approx e^{-iQ j}V_{-}\beta_{+,j}+e^{iQ j}V_{+}\beta_{-,j},
  \\
 & b_{j,2}
 \approx e^{-iQ j}V_{+}\beta_{+,j}+e^{iQ j}V_{-}\beta_{-,j},
 \end{aligned}
 \right.
 \label{eq:b}
\end{equation}
which lead to the following representation of the number operators as 
\begin{equation}
 \left\{
 \begin{aligned}
 n_{j,1}
 & \approx 
 V_{-}^2\tilde{n}_{+,j}
 +V_{+}^2\tilde{n}_{-,j}
 +V_{+}V_{-}
  \left(
   e^{i2Qj}\beta_{+,j}^{\dagger}\beta_{-,j}
   +\mathrm{h.c.}
  \right),
  \\
 n_{j,2}
 & \approx 
 V_{+}^2\tilde{n}_{+,j}
 +V_{-}^2\tilde{n}_{-,j}
 +V_{+}V_{-}
  \left(
   e^{i2Qj}\beta_{+,j}^{\dagger}\beta_{-,j}
   +\mathrm{h.c.}
  \right),
 \end{aligned}
 \right.
 \label{eq:density}
\end{equation}
where the density operators for the separate quadratic energy
dispersions have been defined as
$\tilde{n}_{\pm,j}=\beta_{\pm,j}^{\dagger}\beta_{\pm,j}$.
The above representation of the field operators in the long-wave-length
approximation has a similar form to that of fermions.
Namely the wave numbers $\pm Q$ giving the minima of the energy band are
analogous to Fermi points.

From the above, the Hamiltonian in the long-wave-length-approximation
are derived. 
The single particle Hamiltonian~(\ref{eq:H_0}) is rewritten as 
\begin{equation}
 H_{0}
 \approx
 -E_{0}\sum_{j,\sigma=\pm}\tilde{n}_{\sigma,j}
 +\sum_{q,\sigma=\pm}
  \frac{q^2}{2m^{*}}\beta_{\sigma,q}^{\dagger}\beta_{\sigma,q},
 \label{eq:kinetic--large_flux}
\end{equation}
where $\beta_{\sigma}(q)$ is a Fourier transform of $\beta_{\sigma,j}$.
Using the representation of Eq.~(\ref{eq:density}), the local
Hamiltonian~(\ref{eq:H_U}) is rewritten as 
\begin{align}
 H_{\mathrm{loc}}
 & \approx
 -\left(\mu+\frac{U}{2}-UV_{+}^2V_{-}^2\right)
  \sum_{j,\sigma=\pm}\tilde{n}_{\sigma,j}
 \nonumber \\
 & \quad
 +\frac{U}{4}
  \sum_{j}
  \biggl[
   \left(1+2V_{+}^2V_{-}^{2}\right)
   \left(
    \tilde{n}_{+,j}
    +\tilde{n}_{-,j}
   \right)^2
 \nonumber \\
 & \quad 
   +\left(1-6V_{+}^2V_{-}^2\right)
    \left(
     \tilde{n}_{+,j}
     -\tilde{n}_{-,j}
    \right)^2
 \nonumber \\
 & \quad
   +2V_{+}V_{-}
     \left(\tilde{n}_{+,j}+\tilde{n}_{-,j}\right)
     \left(
      e^{i2Qj}\beta_{+,j}^{\dagger}\beta_{-,j}
      +\mathrm{H.c.}
     \right)
 \nonumber \\
 & \quad
   +2V_{+}V_{-}
     \left(
      e^{i2Qj}\beta_{+,j}^{\dagger}\beta_{-,j}
      +\mathrm{H.c.}
     \right)
     \left(\tilde{n}_{+,j}+\tilde{n}_{-,j}\right)
 \nonumber \\
 & \quad
   +4V_{+}^2V_{-}^2
    \left(
     e^{i4Qj}\beta_{+,j}^{\dagger}\beta_{+,j}^{\dagger}\beta_{-,j}\beta_{-,j}
     +\mathrm{H.c.}
    \right)
  \biggl].
 \label{eq:interaction--large_flux}
\end{align}
The $4Q$-oscillating terms in Eq.~(\ref{eq:interaction--large_flux})
are regarded as the umklapp scattering between the particles in the two
band minima.
Thus the commensurability of the magnetic flux is determined by $Q$.

In order to fix the chemical potential for a given particle number per
site, $\bar{n}_{p}=\langle{n_{j,p}}\rangle$ ($p=1,2$), we use mean-field
analysis.
As discussed in Appendix~\ref{sec:MF}, the mean-field theory leads to
the balanced densities on the chains,
$\bar{n}_{1}=\bar{n}_{2}=\tilde{n}_{+}=\tilde{n}_{-}=\bar{n}$ where
$\tilde{n}_{\pm}=\langle{\tilde{n}_{\pm,j}}\rangle$, 
and the density is controlled by the chemical potential as in
Eq.~(\ref{eq:MF-equation1}). 

Based on the mean-field solution, we use the following bosonization
formula as 
\begin{equation}
 \left\{
  \begin{aligned}
   & \beta_{\pm,j}
   \sim\sqrt{\bar{n}}e^{i\theta_{\pm}(x_{j})},
   \\
   & n_{\pm,j}
   \sim 
   \bar{n}-\frac{a}{\pi}\nabla\varphi_{\pm}(x_j)
   +2\bar{n}\cos\left[\frac{2\pi\bar{n}}{a}x_{j}-2\varphi_{\pm}(x_j)\right],
  \end{aligned}
  \right.
 \label{eq:bosonization1}
\end{equation}
where we have introduced the continuum coordinate $x_j=a \times j$ with
the lattice length $a$.
Note that from the bosonization formula, the fields $\varphi_{\sigma}$
and $\theta_{\sigma}$ are compactified, respectively, as
$\varphi_{\sigma}\sim\varphi_{\sigma}+\pi$ and
$\theta_{\sigma}\sim\theta_{\sigma}+2\pi$.
In other words, the fields are uniquely defined in the regime,
\begin{equation}
 \left\{
 \begin{aligned}
  & -\frac{\pi}{2} < \varphi_{\pm}(x) \le \frac{\pi}{2},
  \\
  & -\pi < \theta_{\pm}(x) \le \pi,
 \end{aligned}
 \right.
\label{eq:compactification1}
\end{equation}
Applying Eq.~(\ref{eq:bosonization1}) into
Eqs.~(\ref{eq:kinetic--large_flux})
and~(\ref{eq:interaction--large_flux}), the low-energy effective
Hamiltonian is derived as 
\begin{align}
 H_{\mathrm{eff}} &
 =H_{\mathrm{TL}}
  +\sum_{i=0}^{4}V_{i},
 \label{eq:effectiveHamiltonian-large_flux}
 \\
 H_{\mathrm{TL}} &
 =\frac{v_\rms}{2\pi}
  \int\!\!dx\,
  \left[
    K_\rms\left(\nabla\theta_\rms(x)\right)^2
    +\frac{1}{K_\rms}\left(\nabla\varphi_\rms(x)\right)^2
  \right]
 \nonumber \\
 &\quad 
 +\frac{v_\rma}{2\pi}
  \int\!\!dx\,
  \left[
    K_\rma\left(\nabla\theta_\rma(x)\right)^2
    +\frac{1}{K_\rma}\left(\nabla\varphi_\rma(x)\right)^2
  \right],
 \label{eq:TLs1}
 \\
 V_{0}
 &=g_{0}\int\!\! \frac{dx}{a}\,
  \cos\left(\frac{2Q}{a}x-\sqrt{2}\theta_\rma(x)\right),
 \label{eq:perturbation0}
 \\
 V_{1}
 &=g_{1}\int\!\! \frac{dx}{a}\,
  \cos\left(\frac{4Q}{a}x-\sqrt{8}\theta_\rma(x)\right),
 \label{eq:perturbation1}
 \\
 V_{2}
 &=g_{2}\int\!\! \frac{dx}{a}\,
   \cos\left(\frac{4\pi\bar{n}}{a}x-\sqrt{8}\varphi_\rms(x)\right),
 \label{eq:perturbation2}
 \\
 V_{3}
 &=g_{3}\int\!\! \frac{dx}{a}\,
   \cos\left(\sqrt{8}\varphi_\rma(x)\right),
 \label{eq:perturbation3}
 \\ 
 V_{4}
 &=g_{4}\int\!\! \frac{dx}{a}\,
   \cos\left(\frac{2\pi\bar{n}}{a}x-\sqrt{2}\varphi_\rms(x)\right)
   \cos\left(\sqrt{2}\varphi_\rma(x)\right),
  \label{eq:perturbation4}
\end{align}
where the symmetric and antisymmetric fields have been introduced as
\begin{align}
 \left\{
  \begin{aligned}
   \varphi_{\rms,\rma}(x) 
   &=\frac{\varphi_{+}(x)\pm\varphi_{-}(x)}{\sqrt{2}},
   \\
   \theta_{\rms,\rma}(x) 
   &=\frac{\theta_{+}(x)\pm\theta_{-}(x)}{\sqrt{2}}.
  \end{aligned}
  \right.
 \label{eq:symmetric-asymmetric_fields}
\end{align}
The Hamiltonian $H_\mathrm{TL}$ stands for that of TL liquids in the
symmetric and antisymmetric sectors.
Note that due to the redefinition of the fields, the compactification
of the fields
changes.~\cite{Oshikawa/arXiv.2010,Oshikawa.Chamon.Affleck/JSTAT.2006,Wong.Affleck/NucPhysB417.1994} 
The redefined fields can not be independently compactified, and the
identification of the fields are as follows: 
$\varphi_{\rms,\rma}\sim\varphi_{\rms,\rma}+\pi N_{\rms,\rma}/\sqrt{2}$
with $N_\rms\equiv N_\rma$ (modulo $2$), and 
$\theta_{\rms,\rma}\sim\theta_{\rms,\rma}+\sqrt{2}\pi M_{\rms,\rma}$ with
$M_\rms\equiv M_\rma$ (modulo $2$).
In other words, the symmetric and antisymmetric fields are uniquely
defined in the following regime, 
\begin{equation}
 \left\{
  \begin{aligned}
   & -\frac{\pi}{\sqrt{2}} 
     < \varphi_\rms(x)\pm\varphi_\rma(x)
     \le \frac{\pi}{\sqrt{2}},
   \\
   & -\sqrt{2}\pi
     < \theta_\rms(x)\pm\theta_\rma(x)
     \le \sqrt{2}\pi,
  \end{aligned} 
  \right.
  \label{eq:compactification2}
\end{equation}
which is important in discussing the degeneracy of the ground states. 
The parameters introduced are roughly estimated as 
\begin{equation}
 \left\{
 \begin{aligned}
  & v_{\rms}
  \sim a\sqrt{\frac{\nbar U}{m^{*}}\left(1+2V_{+}^{2}V_{-}^{2}\right)},
  \\
  & v_{\rma}
  \sim a\sqrt{\frac{\nbar U}{m^{*}}\left(1-6V_{+}^{2}V_{-}^{2}\right)},
  \\
  & K_{\mathrm{s}}
  \sim\pi\sqrt{\frac{\nbar/m^{*}U}{1+2V_{+}^{2}V_{-}^{2}}},
  \\
  & K_{\mathrm{a}}
  \sim\pi\sqrt{\frac{\nbar/m^{*}U}{1-6V_{+}^{2}V_{-}^{2}}},
  \\
  & g_{0}
  \sim 4\nbar^2 UV_{+}V_{-},
  \\
  & g_{1}
  \sim 2\nbar^{2}U V_{+}^{2}V_{-}^{2},
  \\
  & g_{2}
  \sim 8\nbar^{2}UV_{+}^{2}V_{-}^{2},
  \\
  & g_{3}
  \sim 8\nbar^{2}UV_{+}^{2}V_{-}^{2},
  \\
  & g_{4}
  \sim 2\nbar^{2}U\left(1+V_{+}^{2}V_{-}^{2}\right),
 \end{aligned}
 \right.
 \label{eq:parameter_estimation1}
\end{equation}
where
$V_{+}^{2}V_{-}^{2}=\frac{1}{4}\hopratio^2/\sin^2\halfphi\left[\hopratio^2+\sin^2\halfphi\right]$.
The above estimation is valid for a finite but
sufficiently small interaction $U\ll J_{\perp}$ since a small
interaction preserves the nature of the two minima in the single
particle spectrum. 
At stronger coupling, the parameters will be strong influenced by the
renormalization effect due to the irrelevant terms omitted in
Eq.~(\ref{eq:effectiveHamiltonian-large_flux}).
However, the following qualitative tendency of the parameters controlled
by the microscopic parameters is expected to be captured.
In the limit of decoupled chains $J_\perp\rightarrow0$, the
velocities $v_{\rms,\rma}$ and Luttinger parameters $K_{\rms,\rma}$ in
the symmetric and antisymmetric sectors become identical:
$v_\rms/v_\rma\rightarrow 1$, and $K_\rms/K_\rma\rightarrow 1$ as
$J_\perp/J\rightarrow 0$.
For the finite rung hopping $J_\perp$, $K_\rms/K_\rma<1$ and
$v_\rms/v_\rma>1$.
In addition, the velocities and Luttinger parameters are controlled,
respectively, to be enhanced and suppressed by the increase of the
interaction $U$.

It is worthwhile showing the bosonized form of the physical quantity
operators, which is useful when we discuss the physical meaning of the
ordered phases caused by the lock of the fields $\varphi_{\rms,\rma}$ and
$\theta_{\rms,\rma}$. 
The density operators are represented in the bosonized form as
\begin{align}
 n_{j,1}
 &\sim\nbar 
      -\frac{a}{\pi}
       \left[
        V_{-}^{2}\nabla\varphi_{+}(x)+V_{+}^{2}\nabla\varphi_{-}(x)
       \right]
 \nonumber \\
 & \quad
      +2\nbar
       \biggr[
        V_{-}^{2}\cos\left(\frac{2\pi\nbar}{a}x -2\varphi_{+}(x)\right)
 \nonumber \\
 & \quad \qquad \qquad 
        +V_{+}^{2}\cos\left(\frac{2\pi\nbar}{a}x-2\varphi_{-}(x)\right)
       \biggl]
 \nonumber \\
 & \quad
      +2\nbar V_{+}V_{-}
       \cos\left(\frac{2Q}{a}x-\sqrt{2}\theta_{\rma}(x)\right),
 \\
 n_{j,2}
 &\sim\nbar
      -\frac{a}{\pi}
       \left[
        V_{+}^{2}\nabla\varphi_{+}(x)+V_{-}^{2}\nabla\varphi_{-}(x)
       \right]
 \nonumber \\
 & \quad
      +2\nbar
       \biggl[
        V_{+}^{2}\cos\left(\frac{2\pi\nbar}{a}x-2\varphi_{+}(x)\right)
 \nonumber \\
 & \quad \qquad \qquad 
        +V_{-}^{2}\cos\left(\frac{2\pi\nbar}{a}x-2\varphi_{-}(x)\right)
       \biggr]
 \nonumber \\
 & \quad 
       +2\nbar V_{+}V_{-}
        \cos\left(\frac{2Q}{a}x-\sqrt{2}\theta_{\rma}(x)\right).
 \label{eq:density1}
\end{align}
The current operators are defined as
$j^{(\parallel)}_{p,j}=-\partial{H}/\partial{A^{(\parallel)}_{j,p}}$
at the $j$th site on the $p$th chain, and
$j^{\perp}_{j}=-\partial{H}/\partial{A^{(\perp)}_{j}}$ on the $j$th
rung. 
Thus the bosonized form are given as
\begin{align}
 j^{(\parallel)}_{j,1}
 &\sim 2\nbar J
        \biggl[
         V_{+}^2\sin\left(Q-\frac{\phi}{2}\right)
         -V_{-}^2\sin\left(Q+\frac{\phi}{2}\right)
 \nonumber \\
 &\quad
         +aV_{-}^2\cos\left(Q+\frac{\phi}{2}\right)\nabla\theta_{+}(x)
 \nonumber \\
 &\quad
         +aV_{+}^2\cos\left(Q-\frac{\phi}{2}\right)\nabla\theta_{-}(x)
 \nonumber \\
 &\quad
         -2V_{+}V_{-}\sin\halfphi
          \cos\left(\frac{Q}{a}(2x+a)-\sqrt{2}\theta_{\rma}(x)\right)
        \biggr],
 \label{eq:current1-large-flux}\\
 j^{(\parallel)}_{j,2}
 &\sim -2\nbar J
        \biggl[
         V_{+}^2\sin\left(Q-\frac{\phi}{2}\right)
         -V_{-}^2\sin\left(Q+\frac{\phi}{2}\right)
 \nonumber \\
 &\quad
         -aV_{+}^2\cos\left(Q-\frac{\phi}{2}\right)\nabla\theta_{+}(x)
 \nonumber \\
 &\quad
         -aV_{-}^2\cos\left(Q+\frac{\phi}{2}\right)\nabla\theta_{-}(x)
 \nonumber \\
 &\quad
         -2V_{+}V_{-}\sin\halfphi
          \cos\left(\frac{Q}{a}(2x+a)-\sqrt{2}\theta_{\rma}(x)\right)
        \biggr],
 \label{eq:current2-large-flux}\\
 j^{(\perp)}_{j}
 & \sim 2\nbar J_{\perp}\left(V_{+}^2-V_{-}^2\right)
         \sin\left(\frac{2Q}{a}x-\sqrt{2}\theta_{\rma}(x)\right).
 \label{eq:current3-large-flux}
\end{align}
Here in order to somewhat simplify the expression of the density and
current operators, we have mixed the notation of $\varphi_{\pm}$ and
$\theta_{\pm}$ with that of $\varphi_{\rms,\rma}$ and
$\theta_{\rms,\rma}$.
The constant terms of the current operators imply the existence of
Meissner currents, which are non-zero except for $Q\pm\phi/2=\pi N$
($N\in \mathbb{Z}$).

\subsection{Effective Hamiltonian for small magnetic flux}
\label{sec:low-magnetic_flux}
Let us consider the case for a small magnetic flux $\phi<\phi_\mathrm{c}$.
As seen in Figs.~\ref{fig:bands} and~\ref{fig:band_minima}, the lower
single-particle energy band then forms a single minimum at $k=0$, and
the low-energy physics would be governed by the band bottom since the
bosons are expected to dominantly populate the energy minimum.
Thus, similarly to the discussion in
Sec.~\ref{sec:effective_theory_large_flux}, we use the long-wave-length
expansion around the energy minima at $k=0$.
Then the low-energy single-particle spectrum is approximated as
\begin{equation}
 E_{-}(k)
 \approx -E_{0} + \frac{k^2}{2m^{*}},
\end{equation}
where the energy offset and the effective mass have been defined as 
\begin{align}
 & E_{0}
   =J_{\perp}+2J\cos\halfphi,
 \label{eq:dispersion2}
 \\
 & \frac{1}{m^{*}}
   =2J\left[\cos\halfphi-\frac{2J}{J_{\perp}}\sin^2\halfphi\right].
 \label{eq:effective_mass2}
\end{align}
The effective mass~(\ref{eq:effective_mass2}) diverges at the critical
magnetic flux $\phi_{\mathrm{c}}$ given by Eq.~(\ref{eq:critical_flux}),
at which the two band minima merge as in Fig.~\ref{fig:band_minima}.
In such a flux regime near $\phi_{\mathrm{c}}$, we would need higher
orders of $k$ in the approximated dispersion~(\ref{eq:dispersion2}), but
we do not consider such a case in this paper.

We look at the bosonic operators in the long-wave-length approximation.
Projecting out the upper band states, and only considering the small
wave length around the minimum of the lower energy band, i.e., $k=0$,
the bosonic operators~(\ref{eq:u-trans}) are approximated as 
\begin{equation}
 b_{j,1} \approx b_{j,2} \approx \frac{1}{\sqrt{2}}\beta_{j},
 \label{eq:boson_operator2}
\end{equation}
where $\beta_{j}=\frac{1}{\sqrt{N}}\sum_{k}e^{-ik j}\beta(k)$. 
It immediately leads to the approximate form of the density operators as
\begin{equation}
 n_{j,1} \approx n_{j,2} \approx \frac{1}{2}\tilde{n}_{j},
 \label{eq:density_operator2}
\end{equation}
where $\tilde{n}_{j}=\beta^{\dagger}_{j}\beta_{j}$.
This approximate form implies that the densities on the upper and lower
chain are balanced as long as the bosons occupy only the vicinity of the
energy minima.
Using the formulas~(\ref{eq:boson_operator2}) and
(\ref{eq:density_operator2}) in the long-wave-length approximation, the
Hamiltonian~(\ref{eq:H}) is rewritten as
\begin{align}
 H
 &\approx
 \sum_{k}\frac{k^2}{2m^{*}}\beta^{\dagger}_{k}\beta_{k}
 -\left(\mu+E_{0}+\frac{U}{2}\right)\sum_{j}\tilde{n}_{j}
 +\frac{U}{4}\sum_{j}\tilde{n}_{j}^{2}.
 \label{eq:Hamiltonian2}
\end{align}
As in Appendix~\ref{sec:MF}, in this approximation, the chemical
potential to reproduce the density of the original bosons
$\braket{n_{j,1}}=\braket{n_{j,2}}=\nbar$ should be controlled as 
Eq.~(\ref{eq:MF-equation3}), and the corresponding mean density of
$\tilde{n}_j$ is $\langle{\tilde{n}_j}\rangle=2\nbar$.
Based on this mean-field solution, we apply the bosonization,
\begin{equation}
 \left\{
 \begin{aligned}
  & \beta_{j}\sim \sqrt{2\nbar}e^{i\theta(x)},
  \\
  & \tilde{n}_{j} 
    \sim 2\nbar - \frac{a}{\pi}\nabla\varphi(x)
         +4\nbar \cos\left[\frac{4\pi\nbar}{a}x-2\varphi(x)\right].
 \end{aligned} 
 \right.
 \label{eq:bosonization2}
\end{equation}
Then the effective theory of the Hamiltonian~(\ref{eq:Hamiltonian2}), in
which only the fluctuation terms are retained, is straightforwardly
found to be a simple sine-Gordon model: 
\begin{align}
 H_{\mathrm{eff}}
 &=\frac{v}{2\pi}
   \int\!\!dx\, 
    \left[
     K\left(\nabla\theta(x)\right)^2
     +\frac{1}{K}\left(\nabla\varphi(x)\right)^2
    \right]
 \nonumber\\
 &\quad
   +g\int\!\!\frac{dx}{a}\, 
     \cos\left(\frac{4\pi\nbar}{a}x-2\varphi(x)\right),
 \label{eq:effective_theory2}
\end{align}
where the parameters are approximately estimated as
\begin{equation}
 \left\{
 \begin{aligned}
 &v
 \sim a\sqrt{\frac{\nbar U}{m^{*}}},
 \\
 & K
 \sim \pi\sqrt{\frac{\nbar}{m^{*}U}},
 \\
 & g
 \sim 4\nbar^2U.
 \end{aligned}
 \right.
 \label{eq:parameters2}
\end{equation}
The estimation (\ref{eq:parameters2}) applies only at finite but small
$U\ll \max[J_{\perp}, J]$ as mentioned in
Sec.~\ref{sec:effective_theory_large_flux}, but the qualitative tendency
such as an increase and decrease of $v$ and $K$ with $U$, respectively,
is expected to be seen  even if $U$ is not in the limit, as seen in
other cases.
The form of the effective theory~(\ref{eq:effective_theory2}) looks very
similar to that of the one-dimensional Bose-Hubbard
chain~\cite{Giamarchi/2004:book}, but the underlying physics is
different.
To see this, it is useful to look at the bosonized form of the physical
quantities.
The density and current operators of the original bosons are found to be
represented by the bosonization formula~(\ref{eq:bosonization2}) as
\begin{equation}
 \begin{aligned}
 n_{j,1}
 &\sim \nbar-\frac{a}{2\pi}\nabla\varphi(x)
       +2\nbar\cos\left[\frac{4\pi\nbar}{a}x-2\varphi(x)\right],
 \\
 n_{j,2}
 &\sim \nbar-\frac{a}{2\pi}\nabla\varphi(x)
       +2\nbar\cos\left[\frac{4\pi\nbar}{a}x-2\varphi(x)\right],
 \\
 j^{(\parallel)}_{j,1}
 &\sim -\nbar J\sin\halfphi
       +a\nbar J\cos\halfphi\nabla\theta(x),
 \\
 j^{(\parallel)}_{j,2}
 &\sim \nbar J\sin\halfphi
       +a\nbar J\cos\halfphi\nabla\theta(x),
 \\
 j^{(\perp)}_{j}
 &\sim 0. 
 \end{aligned}
 \label{eq:operators-small_flux}
\end{equation}
Note that the currents on the two chains have finite constant terms,
which are proportional to $\sin(\phi/2)$ and have opposite signs, while
the rung current is always zero. 
This implies that for the small magnetic flux $\phi<\phi_\mathrm{c}$,
finite counter-flowing currents are induced on the chains, which
correspond to Meissner currents.

Let us discuss the relation to the argument given in
Ref.~\cite{Orignac.Giamarchi/PRB64.2001} in which a similar problem is
considered, but a different approach is used. 
Orignac and Giamarchi have introduced the independent two phase
fluctuations in the upper and lower chain, i.e.,
$b_{j,p}\propto\exp(i\theta_{j,p})$ for $p=1$, $2$.
In their scenario, the relative phase fluctuation,
$\theta_{j,1}-\theta_{j,2}$, turns out to be gapful because of the
interchain hopping $J_{\perp}$.
On the other hand, in our approach, the higher energy states irrelevant
to the low-energy physics are projected out, which allows us to
effectively identify the bosonic operators, i.e.,
$b_{j,1}\approx b_{j,2}$.
Namely, it means that within our approximation only the in-phase
fluctuation, $\theta_{j,1}+\theta_{j,2}$, is considered as the
phase field here, $\theta(x)$, and the relative phase fluctuation is
omitted in projecting out the higher energy states. Therefore, the
gapful excitation of the relative phase, pointed out by Orignac and
Giamarchi, is associated with the upper band which is projected out in
our treatment.

\section{Discussion}\label{sec:discussion}

We discuss here the ground-state properties based on the obtained effective
theories~(\ref{eq:effectiveHamiltonian-large_flux})
and~(\ref{eq:effective_theory2}).
We consider separately two different limits: 
the case of a large magnetic flux $\phi=\pi$ and the case of a small
magnetic flux $\phi<\phi_\mathrm{c}$. For the latter, the low-energy
single-particle energy band has a single minimum.
In this section, we only consider a filling of one particle per site.

\subsection{Phase diagram for $\pi$ magnetic flux at unity filling}

\subsubsection{General discussion}

The unity filling Bose-Hubbard ladder for a magnetic flux $\phi=\pi$
has been previously discussed by DMRG in
Refs.~\cite{Dhar.etal/PRA85.2012,Dhar.etal/PRB87.2013}, and the
ground-state phase diagram is known to show the following features.
At weak coupling, the system is  in a gapless SF state with staggered
loop currents (chiral superfluid, CSF), while a Mott insulator (MI) is
found in strong-coupling regime.
In between, a MI phase with staggered loop currents (chiral Mott
insulator CMI) is found.
In addition, the criticalities between these phases have also been
numerically studied: the CSF-CMI and CMI-MI transitions exhibit
Berezinskii-Kosterlitz-Thouless 
(BKT)~\cite{Berezinskii/JETP32.1971,Berezinskii/JETP34.1972,Kosterlitz.Thouless/JPhysC6.1973}
and Ising criticality, respectively. 
Here we discuss this ground-state phase diagram from the viewpoint of
the effective field theory.

The momentum giving the energy minima becomes $\pm Q=\pm \pi/2$ 
for $\phi=\pi$ (Fig.~\ref{fig:band_minima}).
In the perturbation $V_{0}$ in the effective theory, an oscillation
remains in the form of 
$\cos\left[\frac{\pi}{2a}x-\sqrt{2}\theta_{\rma}(x)\right]$,
and $V_{0}$ turns out to be irrelevant, while the oscillation in $V_{1}$
is canceled. 
If one considers the second-order perturbation theory in $V_{0}$,
the oscillation cancels:
\begin{equation}
 V_{0}^{2}
 \sim 
 {g'}_{0}
 \int\!\!\frac{dx}{a}\cos\left[\sqrt{8}\theta_{\rma}(x)\right],
 \label{eq:additional-V0}
\end{equation}
where $g'_{0}$ is a coupling constant proportional to $g_{0}^{2}$.
The form of the higher-order contribution~(\ref{eq:additional-V0}) is
identical to that of $V_{1}$, which means that the effect due to $V_{0}$
can be fully absorbed into $V_{1}$.
Thus let us ignore $V_{0}$ in this discussion.
Setting $\nbar=1$, the effective Hamiltonian in the $\pi$ magnetic flux
case turns out to be slightly simplified as
\begin{align}
 H_{\mathrm{eff}}
 &=H_{\mathrm{TL}}
  +\int\!\!\frac{dx}{a}\,
  \biggl[
   g_{1}\cos\left(\sqrt{8}\theta_{\rma}(x)\right)
 \nonumber \\
 &\quad
   +g_{2}\cos\left(\sqrt{8}\varphi_{\rms}(x)\right)
   +g_{3}\cos\left(\sqrt{8}\varphi_{\rma}(x)\right)
 \nonumber \\
 &\quad
   +g_{4}\cos\left(\sqrt{2}\varphi_{\rma}(x)\right)
         \cos\left(\sqrt{2}\varphi_{\rms}(x)\right)
  \biggr],
 \label{eq:effective_Hamiltonian3}
\end{align}
where all the coupling constants are assumed to be positive from the
estimation Eq.~(\ref{eq:parameter_estimation1}).

\begin{figure}
 \centering
 \includegraphics[scale=1.]{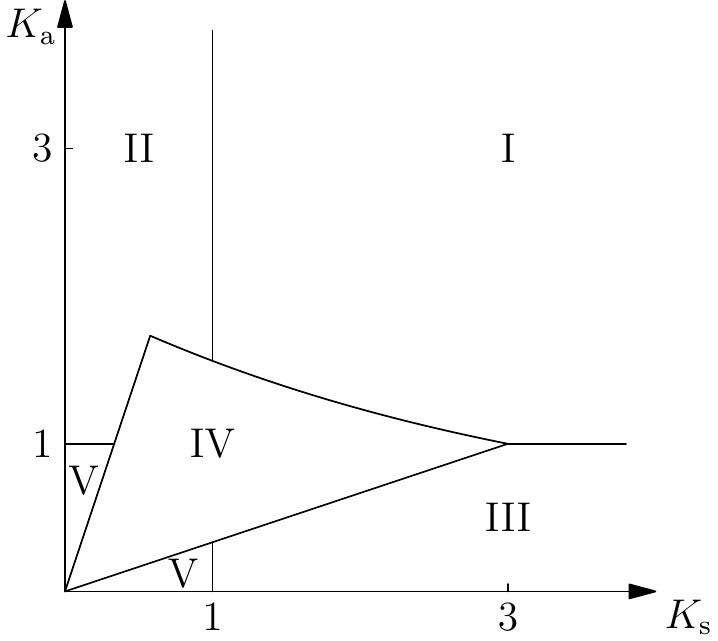}
 \caption{The ground-state phase diagram in the parameter space spanned
 by the Luttinger parameters $K_\rms$ and $K_\rma$.
 The different phases are identified as: chiral superfluid (CSF, Regime I), 
 chiral Mott insulator (CMI, Regime II), conventional superfluid without
 a current pattern (SF, Regime III), conventional Mott insulator without
 a vortex current pattern (MI, Regime IV).
 The phase boundary between regimes I and IV is given by
 $K_\rma=-K_\rms/2+\sqrt{(K_\rms/2)^2+4}$, and the one between IV and V
 by $K_\rma=3K_\rms$ and $K_\rma=K_\rms/3$.}
 \label{fig:phase_diagram}
\end{figure}

The derived effective theory~(\ref{eq:effective_Hamiltonian3}) is still
complicated to analyze.
We thus discuss the possible phases from the viewpoint of a scaling analysis.
Let us consider a perturbative renormalization-group treatment of all
the cosine terms in the effective
Hamiltonian~(\ref{eq:effective_Hamiltonian3}), and identify the scaling
dimension of those cosine terms around the Gaussian fixed point.
Denoting by $x_{\mathcal{O}}$ the scaling dimension of a perturbation
$\mathcal{O}$, we obtain for the effective
theory~(\ref{eq:effective_Hamiltonian3}) the following values: 
\begin{equation}
 \left\{
  \begin{aligned}
   & x_{\cos(\sqrt{8}\theta_{\rma})}=\frac{2}{K_{\rma}},
   \\
   & x_{\cos(\sqrt{8}\varphi_{\rms})}=2K_{\rms},
   \\
   & x_{\cos(\sqrt{8}\varphi_{\rma})}=2K_{\rma},
   \\
   & x_{\cos(\sqrt{2}\varphi_{\rma})\cos(\sqrt{2}\varphi_{\rms})}
     =\frac{K_{\rms}+K_{\rma}}{2}.
  \end{aligned}
 \right.
\end{equation}
Up to first-order perturbative renormalization group, relevant
perturbations $\mathcal{O}$ are those for which $x_{\mathcal{O}}<2$.
To derive the effective field theory depending on the possible values of
the Luttinger parameters, we take the following steps:
\begin{enumerate}
 \item First the possible relevant terms, whose scaling dimensions are
       $<2$, are written down depending on the parameter regime of
       $K_\rms$ and $K_\rma$. 
 \item Referring to the relevancy of the perturbations, we divide the
       parameter space into several subspaces.
       In each subspace the low-energy physics is described by an
       effective theory consisting of a different set of relevant
       perturbations.
 \item If there are several relevant perturbations in the subspace, each
       perturbation tends to lock the fields of  
       $\varphi_{\rms,\rma}$ and $\theta_{\rms,\rma}$ to be different
       values. 
       Then, if some of the relevant perturbations compete so as to fix
       the fields to the different values, e.g., the pairs of 
       $\cos(\sqrt{8}\theta_\rma)$ and $\cos(\sqrt{8}\varphi_\rma)$, and
       of $\cos(\sqrt{8}\varphi_\rms)$ and
       $\cos(\sqrt{2}\varphi_\rms)\cos(\sqrt{2}\varphi_\rma)$, we 
       retain only the most relevant perturbation, and omit the
       competing less relevant ones.
 \item If some of the relevant terms do not compete, e.g.,
       $\cos(\sqrt{8}\theta_{\rma})$ and $\cos(\sqrt{8}\varphi_{\rms})$,
       we retain all of them.
\end{enumerate}
Following this procedure, the parameter space spanned by the 
Luttinger parameters $K_\rms$ and $K_\rma$ is found to be separated into
five different regimes, as displayed on Fig.~\ref{fig:phase_diagram},
and each regime is governed by a particular form of the low-energy
effective theory.

\begin{figure}
 \centering
 \includegraphics[scale=1.]{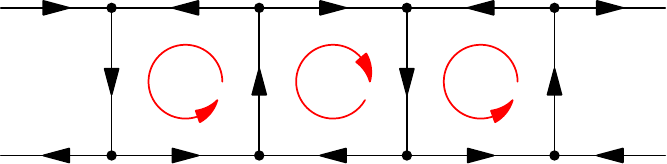}
 \caption{(Color online) A current pattern associated with the CSF phase
 (Regime I) and CMI phase (Regime II) for a $\pi$ magnetic flux per
 plaquette. 
 The black arrows denote the local currents given by
 Eq.~(\ref{eq:current-pattern}).
 The red circular arrows denote the local staggered vortices deduced
 from the local current pattern.
 The local currents vanish at small $J_\perp/J$, and their strength
 increases with the rung hopping $J_\perp/J$.
 }
 \label{fig:current-pattern}
\end{figure}

In Regime I, $K_\rms>1$, $K_\rma>1$ and
$K_\rma>-K_\rms/2+\sqrt{(K_\rms/2)^2+4}$, the low-energy effective 
theory is given as 
\begin{align}
 H^{\mathrm{(I)}}_{\mathrm{eff}}
 =H_{\mathrm{TL}}
  +g_1\int\!\!\frac{dx}{a}\cos\left(\sqrt{8}\theta_\rma(x)\right).
 \label{eq:effective-theoryI}
\end{align}
Due to the cosine term, the relative phase $\theta_\rma$ is locked in
the ground states as $\langle{\theta_\rma}\rangle=\pm\pi/\sqrt{8}$,
which generates the finite energy gap in the antisymmetric field sector,
while the unbounded symmetric phase sector remains gapless.
According to the bosonized form of the current
operators~(\ref{eq:current1-large-flux})-(\ref{eq:current3-large-flux}),
the lock of the field $\theta_\rma$ leads to the local currents: 
In the case of the fixed relative phase
$\langle{\theta_\rma}\rangle=\pi/\sqrt{8}$, 
\begin{equation}
 \left\{
  \begin{aligned}
   & \langle{j^{(\parallel)}_{j,1}}\rangle \sim -4\nbar J V_{+}V_{-}(-1)^{j},
   \\
   & \langle{j^{(\parallel)}_{j,2}}\rangle \sim 4\nbar J V_{+}V_{-}(-1)^{j},
   \\
   & \langle{j^{(\perp)}_{j}}\rangle \sim -2\nbar J_{\perp}(V_{+}^2-V_{-}^2)(-1)^{j},
  \end{aligned}
 \right.
 \label{eq:current-pattern}
\end{equation}
and for $\langle{\theta_\rma}\rangle=-\pi/\sqrt{8}$ the sign of all the
currents becomes opposite. 
Because
$V_{+}V_{-}=\frac{1}{2}(J_\perp/2J)^2/\sqrt{(J_\perp/2J)^2+1}$ and
$J_\perp(V_{+}^2-V_{-}^2)=J_\perp/\sqrt{1+(J_\perp/2J)^2}$ for
$\phi=\pi$, the currents in Eq.~(\ref{eq:current-pattern}) disappear at
the limit of $J_\perp/J\rightarrow 0$, and the strength grows as
$J_\perp/J$ goes up.
The currents on the $j$th bond in the upper and lower chain point
oppositely, and the rung current are staggered along the chain
direction. 
Based on the representation~(\ref{eq:current-pattern}), the current
pattern is illustrated on Fig.~\ref{fig:current-pattern}, in which 
staggered loop currents are found to appear. 
Therefore, Regime I should be interpreted as a CSF phase. 
The two-fold degeneracy is caused by the spontaneous breaking of
translation symmetry~\footnote{Note that the choice of
gauge~(\ref{eq:gauge_choice}) preserves translational symmetry of the
Hamiltonian. If we take another choice of gauge,
$A^{\parallel}_{j,1}=A^{\parallel}_{j,2}=0$ and
$A^{\perp}_{j}=-\phi\times j$, the translation symmetry preserved in the
choice of gauge~(\ref{eq:gauge_choice}) is explicitly broken.}.
This physical description of the CSF phase agrees with that given in
Refs.~\cite{Dhar.etal/PRA85.2012,Dhar.etal/PRB87.2013}.

In Regime II, $K_\rms<1$, $K_\rma>1$, $K_\rma>3K_\rms$ and
$K_\rma>-K_\rms/2+\sqrt{(K_\rma/2)^2+4}$, the low-energy effective
theory is given as
\begin{align}
 H^{\mathrm{(II)}}_{\mathrm{eff}}
 &=H_{\mathrm{TL}}
  +\int\!\!\frac{dx}{a}
   \biggl[
    g_1\cos\left(\sqrt{8}\theta_\rma(x)\right)
    +g_2\cos\left(\sqrt{8}\varphi_\rms(x)\right)
   \biggr].
 \label{eq:effective-theoryII}
\end{align}
The two cosine terms in the effective theory separately lock both the
relative phase $\theta_\rma$ and the symmetric field $\varphi_\rms$ in
the ground state:
$\langle{\theta_\rma}\rangle=\pm\pi/\sqrt{8}$ and
$\langle{\varphi_\rms}\rangle=\pm\pi/\sqrt{8}$.
The two locked values $\langle{\varphi_\rms}\rangle=\pm\pi/\sqrt{8}$
cannot be distinguished by the compactification
condition~(\ref{eq:compactification2}), but do not lead to any
difference in the physical
quantities~(\ref{eq:density1})-(\ref{eq:current3-large-flux}). 
Thus we can identify the two locks
$\langle{\varphi_\rms}\rangle=\pm\pi/\sqrt{8}$ from the physical
viewpoint, and in total the ground states are found to be two-fold
degenerate.
Due to the locking of the two fields, an energy gap opens both in
the symmetric and antisymmetric sector, and thus the low-energy
excitations in Regime II are fully gapped.
As discussed in Regime I, the values of the locked relative phase,
$\langle{\theta_\rma}\rangle=\pm\pi/\sqrt{8}$, result in the current
pattern (\ref{eq:current-pattern}) illustrated by
Fig.~\ref{fig:current-pattern}.
On the other hand, the locking of the field $\varphi_\rms$ physically
means that the density fluctuation is frozen, which means that the
system behaves like a MI. 
Thus Regime II should be identified with the CMI phase which involves
the current pattern shown in Fig.~\ref{fig:current-pattern}.
From this current pattern, we can physically expect a two-fold
degeneracy of the ground state, and this degeneracy comes from the two
possible locks of $\theta_\rma$.

In Regime III, $K_\rms>1$, $K_\rma<1$ and $K_\rma<K_\rms/3$, the
low-energy effective theory is given by
\begin{align}
 H^{\mathrm{(III)}}_{\mathrm{eff}}
 =H_{\mathrm{TL}}
  +g_3\int\!\!\frac{dx}{a}\cos\left(\sqrt{8}\varphi_\rma(x)\right).
 \label{eq:effective-theoryIII}
\end{align}
The antisymmetric field $\varphi_\rma$ is fixed in the ground
state, i.e., $\langle{\varphi_\rma}\rangle=\pm\pi/\sqrt{8}$, and the
excitation in this antisymmetric sector becomes gapful, 
while the symmetric sector remains gapless. 
The physical meaning of this lock of the field $\varphi_\rma$ is not
clear because both the density and currents do not show a signature of
the corresponding order.
The two possible locks of $\varphi_\rma$ result in the double degeneracy
of the ground states, but they do not give any difference in the
physical quantities~(\ref{eq:density1})-(\ref{eq:current3-large-flux}).
From the above, we can conclude that the ground state in Regime III is 
unique and some kind of SF phase with one gapless excitation mode.

In Regime IV, $K_\rma>K_\rms/3$, $K_\rma<3K_\rms$ and
$K_\rma<-K_\rms/2+\sqrt{(K_\rma/2)^2+4}$, the low-energy effective
theory is given as 
\begin{align}
 H^{(\mathrm{IV})}_{\mathrm{eff}}
 &=H_{\mathrm{TL}}
  +g_4\int\!\!\frac{dx}{a}
       \cos\left(\sqrt{2}\varphi_\rms(x)\right)
       \cos\left(\sqrt{2}\varphi_\rma(x)\right).
 \label{eq:effective-theoryIV}
\end{align}
Thus in the ground state both the fields $\varphi_\rma$ and
$\varphi_\rms$ are naively found to be locked in the following two ways:
$\langle{\varphi_\rma}\rangle=\pi/\sqrt{2}$ 
and $\langle{\varphi_\rms}\rangle=0$,
or
$\langle{\varphi_\rma}\rangle=0$ 
and $\langle{\varphi_\rms}\rangle=\pi/\sqrt{2}$.
However, from the compactification~(\ref{eq:compactification2}), these
two locked points are identical, and thus the ground state is unique.
Due to the locking of the two fields $\varphi_\rms$ and $\varphi_\rma$,
which means that all the density fluctuations are frozen, the ground
state is fully gapped.
Therefore Regime IV corresponds to the conventional MI phase. 

In Regime V, $K_\rms<1$ and $K_\rma<K_\rms/3$ or
$K_\rma<1$ and $K_\rma>3K_\rms$, the low-energy effective
theory is given as 
\begin{align}
 H^{(\mathrm{V})}_{\mathrm{eff}}
 &=H_{\mathrm{TL}}
  +\int\!\!\frac{dx}{a}
   \biggl[
    g_{2}\cos\left(\sqrt{8}\varphi_\rms(x)\right)
    +g_{3}\cos\left(\sqrt{8}\varphi_\rma(x)\right)
   \biggr].
 \label{eq:effective-theoryV}
\end{align}
Thus in the ground state, $\varphi_\rms$ and $\varphi_\rma$ are fixed to
be $\langle{\varphi_\rms}\rangle=\pm\pi/\sqrt{8}$ and 
$\langle{\varphi_\rma}\rangle=\pm\pi/\sqrt{8}$.
However, due to the compactification~(\ref{eq:compactification2}),
$(\langle{\varphi_\rms}\rangle,\langle{\varphi_\rma}\rangle)=(-\pi/\sqrt{8},\pm\pi/\sqrt{8})$
are identified with $(\pi/\sqrt{8},\mp\pi/\sqrt{8})$, respectively. 
Thus the two distinguishable states minimize the cosine terms in the
effective theory. 
In order to clarify the physical meaning of these ground states in this
phase, we look at the bosonized form of the density difference between
the chains from Eq.~(\ref{eq:density1}).
Then the mean values of the density difference is found to give 
\begin{equation}
 \langle{n_{j,1}-n_{j,2}}\rangle
 \sim
 4\left(V_{+}^{2}-V_{-}^{2}\right)
 \sin\left(\sqrt{2}\langle{\varphi_\rms}\rangle\right)
 \sin\left(\sqrt{2}\langle{\varphi_\rma}\rangle\right).
\end{equation}
The density difference is found to be finite in the obtained two states:
$\langle{n_{j,1}-n_{j,2}}\rangle>0$ for
$\langle{\varphi_\rms}\rangle=\pi/\sqrt{8}$ and 
$\langle{\varphi_\rma}\rangle=\pi/\sqrt{8}$,
and 
$\langle{n_{j,1}-n_{j,2}}\rangle<0$ for
$\langle{\varphi_\rms}\rangle=\pi/\sqrt{8}$ and 
$\langle{\varphi_\rma}\rangle=-\pi/\sqrt{8}$.
Such a density imbalance is inconsistent with the balanced density
situation based on the mean-field analysis in Appendix~\ref{sec:MF}.
Thus the simultaneous lock of both fields $\varphi_\rms$ and
$\varphi_\rma$ signals the instability of the state with balanced
densities between the two chains, leading to a state with density
imbalance (DI).

\subsubsection{Physical phase diagram as a function of 
$U/J$ and $J_\perp/J$}

We have discussed the general structure of the phase diagram
(Fig.~\ref{fig:phase_diagram}), but the SF and DI phase
may not be realized in the original Bose-Hubbard model due to 
the two following reasons.
The first is that the regime $K_\rma < K_\rms$ would be forbidden 
in terms of the microscopic parameters $(U/J,J_\perp/J)$.
This is predicted by the naive parameter
estimation~(\ref{eq:parameter_estimation1}).
Thus the SF phase (Regime III) and a part of the DI phase (Regime V)
would not be realistic.
The other reason is that the Luttinger parameters in these regimes would
be too small to reach.
Naively a Luttinger parameter for bosons with short-range interaction
such as the Lieb-Liniger model~\cite{Cazalilla/JPhysB37.2004} and
the Bose-Hubbard model at an incommensurate
filling~\cite{Giamarchi/2004:book} can run only from 
infinity to unity as interaction increases, in which the infinite and
unity limit of the Luttinger parameter correspond to the non-interacting
and hard-core boson limit, respectively.
Strictly speaking, these constraints do not necessarily apply the
present ladder model, but $K_\rms<1/3$ or $K_\rma<1/3$ for the DI phase
(Regime V) is still considered to be extremely small for a bosonic
system.
Indeed the numerically determined phase diagram given in
Refs.~\cite{Dhar.etal/PRA85.2012,Dhar.etal/PRB87.2013} does not show
such SF and DI phases.

The obtained phase diagram Fig.~\ref{fig:phase_diagram} is parametrized
by the phenomenological parameters $K_\rms$ and $K_\rma$.
Thus in order to estimate the phase diagram in terms of the microscopic
parameters, $U/J$ and $J_\perp/J$, we need to clarify the behavior of
the Luttinger parameters as a function of these microscopic parameters.
The general field theory analysis, which applies only at low energy, is
insufficient to fully answer to this microscopic question. 
Thus we make use of other general arguments and constraints to figure
out qualitatively the phase diagram in terms of the microscopic
parameters. 

\begin{figure}
 \centering
 \includegraphics[width=\plotwi]{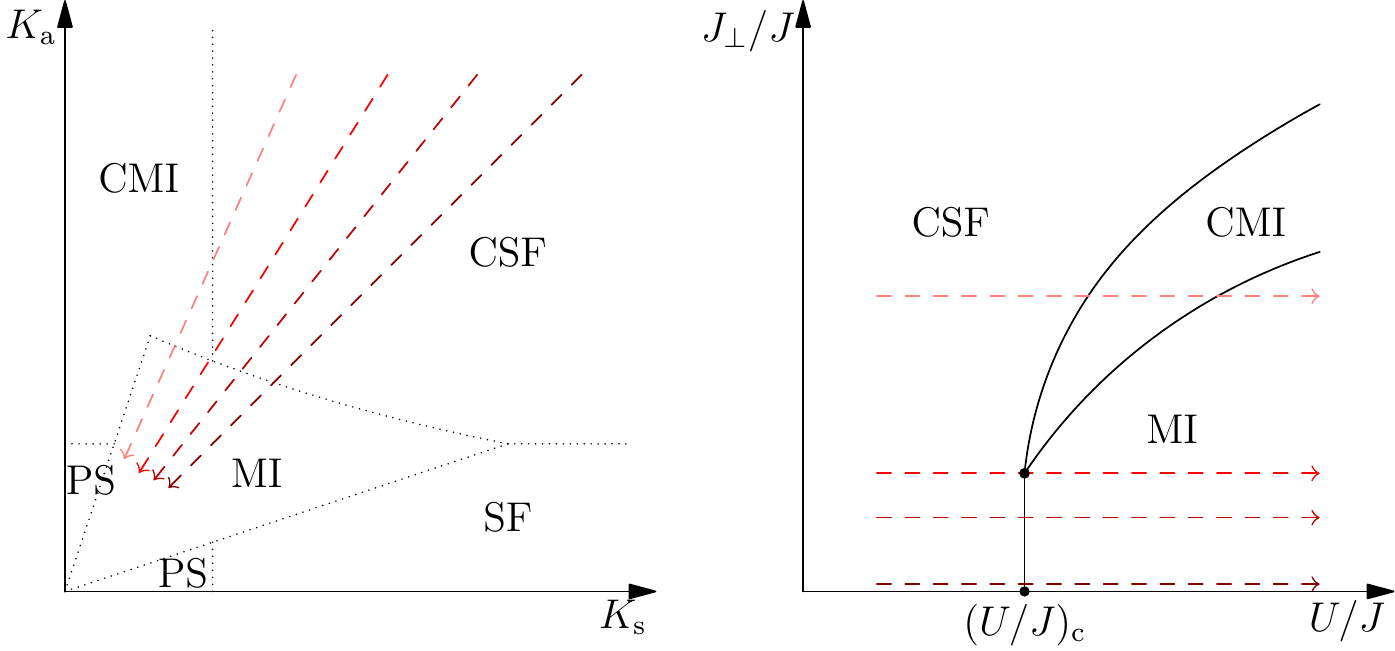}
 \caption{(Color online) A sketch of the trajectories of $K_\rms$ and
 $K_\rma$ for different values of $J_\perp/J$ (left panel), and
 schematic phase diagram as a function of the microscopic parameters
 (right panel).
 Each arrow in the left panel indicates the evolution of  $K_\rms$ and
 $K_\rma$ as the interaction strength $U/J$ is increased.
 The rightmost arrow whose trajectory is described by 
 $K_\rms\simeq K_\rma$ corresponds to the limit of decoupled chains
 $J_\perp/J\rightarrow 0$, and other arrows correspond to the gradual
 increase of $J_\perp/J$.
 Following the different trajectories in the generic phase diagram 
 allows one to establish qualitatively the physical phase diagram
 displayed in the right panel.
 }
 \label{fig:translation_phase_diagram}
\end{figure}

The following qualitative features of the Luttinger parameters can be
deduced from the estimation in Eq.~(\ref{eq:parameter_estimation1}). 
The Luttinger parameter of the symmetric sector is smaller than
that of the antisymmetric sector for given $U/J$ and $J_\perp/J$, i.e.,
$K_\rms < K_\rma$. 
In addition, $K_\rms/K_\rma\rightarrow 1$ as the ladder is decoupled
$J_\perp/2J\rightarrow0$. 
The Luttinger parameters must be large at small interaction $U$ and
decrease as the interaction goes stronger.
From these assumptions, we can expect the following evolution of the
trajectory between $K_\rms$ and $K_\rma$ by controlling the interaction
$U$: At the limit $J_\perp/2J\rightarrow 0$, $K_\rma=K_\rms$, and this
trajectory continuously deforms keeping $K_\rma>K_\rms$ as $J_\perp/2J$
grows.
The expected trajectories are shown in the left panel of
Fig.~\ref{fig:translation_phase_diagram}.

As clear from Fig.~\ref{fig:translation_phase_diagram}, the trajectory
$K_\rma=K_\rms$ implies that the system is in the CSF phase in the
weakly interacting regime, and becomes MI at a critical value of $U/J$ 
without an intervening CMI phase. 
Since the deformation of the trajectory by a change of $J_\perp/J$
should be continuous, the above SF-MI transition must remain up to a
certain value of $J_\perp/J$.
At a specific value of $J_\perp/J$, the trajectory passes the
tricritical point at which the phase boundaries among the CSF (Regime
I), CMI (Regime II), and MI (Regime IV) phase meet.
Beyond this value of $J_\perp/J$, a CMI phase (Regime II) opens up in
between the CSF and MI phases for intermediate interaction strengths
$U/J$.
We summarize this description and the deduced phase diagram in the space
of microscopic parameters in Fig.~\ref{fig:translation_phase_diagram}.

The important point of the deduced phase diagram
Fig.~\ref{fig:translation_phase_diagram} is the presence of the
tricritical point.
In the DMRG study of Ref.~\cite{Dhar.etal/PRA85.2012,Dhar.etal/PRB87.2013} 
this tricritical point was not found, presumably because of the limited
number of values of the coupling constants that were investigated. 
The absence of the CMI phase for small $J_\perp/J$ can also be
established from another field-theoretical approach.
As in Ref.~\cite{Orignac.Giamarchi/PRB64.2001}, if we bosonize the
Hamiltonian~(\ref{eq:H}) in the limit of $J_\perp/J=0$, and take into
account the rung hopping perturbatively, the first-order contribution of
the rung hopping Hamiltonian involves a $\pi$-oscillating term, and thus 
we need to take into account at least the second-order perturbation in
order to see the finite rung hopping contribution.
It means that for sufficiently small rung hopping, the Bose-Hubbard
ladder in the presence of a magnetic flux can be effectively identified
to decoupled Bose-Hubbard chains.
Thus, in such a small rung hopping regime, one can only observe the
SF-MI transition by controlling the interaction $U/J$ as in the case of
the single Bose-Hubbard chain.
In addition, from this argument, the SF-MI transition line drawn by
controlling $J_\perp/J$ is inferred to be independent of $J_\perp/J$.
Namely the boundary between the CSF and MI phase rises up from
$J_\perp/J=0$ perpendicularly to the $U/J$ axis, and eventually 
bifurcates into the two lines of the CSF-CMI and CMI-MI transitions.

\subsubsection{Critical behavior}

Closing the discussion on the ground-state phase diagram of the $\pi$
magnetic flux case at unity filling, we discuss the nature of the 
quantum critical behavior between the different phases. 

Let us first consider the CSF-CMI transition.
As in the effective theories, Eq.~(\ref{eq:effective-theoryI}) for CSF,
and Eq.~(\ref{eq:effective-theoryII}) for CMI, the symmetric and
antisymmetric sectors are decoupled in both regimes, and the transition
is found to be characterized by the locking of the symmetric field
$\varphi_\rms$. 
Hence, focusing only on the symmetric sector in these two regimes, the
phase transition from CSF to CMI is analogous to that of the sine-Gordon
model.
Thus the CSF-CMI transition is concluded to be of BKT transition nature,
which is in agreement with the statement made in the numerical study of 
Ref.~\cite{Dhar.etal/PRA85.2012,Dhar.etal/PRB87.2013}.

The nature of the CMI-MI transition is a more complicated issue, 
because the symmetric and antisymmetric sectors are coupled in the
effective theory~(\ref{eq:effective-theoryIV}) of the MI regime. 
Comparing the effective theories $H^{(\mathrm{II})}_{\mathrm{eff}}$ and
$H^{(\mathrm{IV})}_{\mathrm{eff}}$, two phenomena are found to occur 
at the transition from CMI to MI. 
One is the switch of the locked field in the antisymmetric sector, 
from $\theta_\rma$ to $\varphi_\rma$, and the other is the change of the
locking value of the symmetric field $\varphi_\rms$.
In addition, the two-fold degeneracy caused by the fixed $\theta_\rma$
in the CMI phase is found to change to a non-degenerate state in the MI
phase. 
This change of the degeneracy means that the translation symmetry which
is spontaneously broken in the CMI phase is restored in the MI phase. 
The nature is thus analogous to the $\mathbb{Z}_{2}$ Ising transition,
which was also predicted for the CMI-MI transition for the frustrated
bosonic ladder system~\cite{Zaletel.et.al/PRB89.2014}. 
The previous numerical
studies~\cite{Dhar.etal/PRA85.2012,Dhar.etal/PRB87.2013} has pointed out
the $\mathbb{Z}_{2}$ Ising criticality of the CMI-MI transition, and our 
field theoretical approach is thus consistent with this. 

Finally we consider the direct phase transition between CSF and MI, 
shown in Fig.~\ref{fig:translation_phase_diagram}. 
Because of the coupling of the symmetric and antisymmetric sector in the
MI phase, the analysis of this transition is not simple. 
Two simultaneous phenomena occur: the switch of the bound field from
$\theta_\rma$ to $\varphi_\rma$ in the asymmetric sector and the locking
of the field $\varphi_\rms$ in the symmetric sector. 
This phase boundary is intriguing because two different symmetries are
simultaneously involved: the continuous $O(2)$ symmetry associated with
the SF and the $\mathbb{Z}_{2}$ symmetry associated with the breaking of
translational invariance in the CSF phase.
From usual considerations based on the Landau-Ginzburg-Wilson approach
to critical phenomena, one may conclude that this phase transition is
first-order.
Indeed, according to Eq.~(\ref{eq:current-pattern}), the local currents
in the CSF phase do not depend on the interaction $U$, and the loop
current is thus expected to discontinuously vanish when crossing the
phase boundary from the CSF to MI phase.
However, because the discussion which leads to the loop
currents~(\ref{eq:current-pattern}) is a kind of mean-field approach, a
more in-depth discussion would be needed to obtain a crucial conclusion
on the criticality.
A more intriguing possibility on this criticality is that it might
nonetheless be second-order, despite breaking simultaenously two
unrelated
symmetries~\cite{Senthil.et.al/Science303.2004,Senthil.et.al/PRB70.2004}.

\subsection{The ground state for small magnetic flux at unity filling}

Next we discuss the phase diagram of the Bose-Hubbard ladder for a
sufficiently small magnetic flux in which the bottom of the single
particle spectrum shows a single energy minimum structure. 
In addition, we fix the filling at one particle per site. 
Then, setting $\nbar=1$, we can write the effective
Hamiltonian~(\ref{eq:effective_theory2}) as 
\begin{align}
 H_{\mathrm{eff}}
 &=\frac{v}{2\pi}
   \int\!\!dx\,
    \left[
     K\left(\nabla\theta(x)\right)^2
     +\frac{1}{K}\left(\nabla\varphi(x)\right)^2
    \right]
 \nonumber \\
 &\quad
   +g\int\!\!\frac{dx}{a}\,\cos\left(2\varphi(x)\right).
 \label{eq:SG_model}
\end{align}
It has the same form as that of the single Bose-Hubbard chain.
Thus a BKT transition is found when the Luttinger parameter as defined
here reaches $K=2$. This transition is identical to the SF-Mott
insulator
transition~\cite{Giamarchi/2004:book,Cazalilla.et.al/RMP83.2011}. 
In the weakly interacting regime, the Luttinger parameter is
larger than the critical value $K=2$, and the system is a gapless
TL liquid, i.e. a one-dimensional SF.
As the interaction is tuned to be larger, the Luttinger parameter
becomes smaller, and the system becomes a MI for $K<2$.
This behavior can be captured by the approximately estimated
parameters~(\ref{eq:parameters2}): $K\propto U^{-1/2}$.
The Luttinger parameter should be determined in terms of the interaction
$U/J$ and the rung hopping $J_\perp/J$, but the corresponding critical
value of these microscopic parameters can not be determined just from
the field theoretical argument. 
Thus in this paper we do not discuss further quantitatively the
ground-state phase diagram of the effective theory~(\ref{eq:SG_model})
in the microscopic parameter space. 

\begin{figure}
 \centering
 \includegraphics[scale=1.]{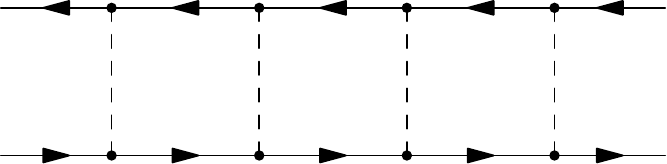}
 \caption{A pattern of Meissner currents appearing for a small magnetic
 flux. The strength of the Meissner current increases with the magnetic
 flux $\phi$.
 This current also remains even in the Mott insulator phase, but the
 chiral state is not gapless.
 }
 \label{fig:meissner}
\end{figure}

Let us look at the ground-state physical properties of the gapless SF
and MI phase predicted by the effective theory~(\ref{eq:SG_model}). 
As mentioned in Sec.~\ref{sec:low-magnetic_flux}, the bosonized form of
the current operators in Eq.~(\ref{eq:operators-small_flux}) implies a 
non-zero constant current, which is displayed in
Fig.~\ref{fig:meissner}.
As given in the form, $\pm J\sin(\phi/2)$, this persistent current
is induced by the magnetic flux, and is thus interpreted to be a
Meissner current in the case of the ladder geometry.
What is interesting is that the presence of this Meissner current is
independent of the physics of the density fluctuation $\varphi$. 
On the other hand, even when the system is in the MI phase, in which
$\varphi$ is locked by the cosine term, the Meissner current remains. 

The physical reason of the presence of the Meissner current in the MI
phase can be understood as follows.
As well known, in the MI state, the phase of each bosons is completely
disordered since the canonically conjugate density fluctuations are
frozen.
However, this statement does not forbid the lock of the relative phase
between the bosons of the different component. 
Thus, in this MI case, each phase of the bosons on the upper and on the
lower chain is disordered, but the relative phase between them is kept to
be locked like that of the SF phase.
Indeed, as mentioned in Ref.\cite{Orignac.Giamarchi/PRB64.2001}, the
Meissner current is a consequence of the lock of the relative phase
between the bosons on the upper and on the lower chain.
A similar nature of the Meissner current in the fully gapped ground
state was also pointed out in Ref.~\cite{Petrescu.LeHur/PRL111.2013}.

A question which naturally arises is what happens to the MI with
Meissner currents in the limit of strong interactions.
In our effective field theory approach~(\ref{eq:SG_model}), only two
phases (SF and MI with Meissner currents) have been obtained. 
However, if we turn back to the original microscopic Hamiltonian, we do
expect the current-carrying Mott state to be eventually unstable in
favor of a conventional Mott state without currents. 
Indeed, in the weak-coupling effective field-theory, one first
establishes the two-band structure of the non-interacting Hamiltonian
and then turns on an interaction within the lowest band only (protected
by a gap from the upper one). 
For a strong interaction, however, matrix elements of the interaction
will couple the two bands, which may break the relative phase coherence
between the two chains and lead to a conventional MI. 
This regime is away from the range of applicability of the effective
field-theory approach.

From the above, we can describe the ground state of the unity-filling
Bose-Hubbard ladder at a small magnetic flux as follows.
In the weakly interacting regime, the system is a SF with Meissner
currents, and the low-energy excitations carry chiral current, i.e.,
the directions of the carried currents on the upper and on the lower
chain are opposite each other.
On the other hand, in the strong interaction regime, the system is a MI
in which the density fluctuations are completely suppressed, and there
are no gapless excitations.
However, this MI state still includes the Meissner current background.
The transition between these two SF and MI phase is a BKT transition,
as seen in a simple one-dimensional Bose-Hubbard chain at an integer
filling.
Furthermore, at strong interaction $U\gg J, J_\perp$, the MI with the
Meissner current should turn into a conventional MI without currents.
However, the transition and criticality between these two MI phases are
not easily addressed within the present analysis.

\section{Summary and perspectives}\label{sec:summary}

In this paper, we have discussed the Bose-Hubbard model with a uniform
magnetic flux in ladder geometry.
Discussing the small and large magnetic flux limits separately, we have
constructed in each case the appropriate low-energy effective field
theory by using bosonization techniques based on the nature of the
single-particle spectrum.
The key difference between the two cases is the number of lowest energy
band minima. 
For a small magnetic flux, the bottom of the lowest band displays a single
minimum. 
Increasing the magnetic flux beyond a critical value $\phi_\mathrm{c}$,
this single minimum splits into two degenerate minima, which leads
to a different structure of the low-energy field theory.

As an application of the derived effective field theories, we have
discussed in detail the phases and physical properties of the system
with one particle per site in the two cases of $\phi=\pi$ and small
$\phi<\phi_\mathrm{c}$.
For the $\pi$ magnetic flux, we have established the general
ground-state phase diagram as a function of the Luttinger parameters
characterizing the low-energy field theory.
Several phases appear: a superfluid (SF), chiral superfluid (CSF), 
Mott insulator (MI), chiral Mott insulator (CMI), as well as a regime of
density imbalance (DI).
Furthermore, we have also discussed the mapping of this generic phase
diagram in terms of the two microscopic parameters of the Bose-Hubbard
model (the interaction strength $U/J$ and ratio of rung to in-chain
hopping $J_\perp/J$).
We have established that the CMI phase only occurs beyond a critical
value of $J_\perp/J$, and to reveal the existence of a tricritical point
at which the CSF, CMI and MI phases meet together.
We also discussed the zero-temperature transitions and critical behavior
separating these phases, and pointed out that the precise nature of the
critical behavior for the direct transition between the CSF and MI phase
is an interesting open issue to be addressed in future studies.

In the small magnetic flux case, we have clarified the possible ground
states and their properties.
The SF and MI states have been, respectively, found to appear at weak
and strong interaction strength, with a BKT transition between them.
We found that not only the SF state but also the MI state displays
Meissner currents.

In a remarkable recent experiment~\cite{Atala.et.al/arXiv2014}, Atala
{\it et al.} realized a two-leg ladder optical lattice in which bosonic
atoms are confined and subject to an artificial uniform magnetic field. 
The Meissner currents and vortex currents in the SF phase were
successfully probed by using a site-resolved local current 
measurement~\cite{Trotzky.et.al/NatPhys8.2012,Kessler.Marquardt/arXiv2013}. 
These achievements should make it possible to investigate experimentally
the various phases (SF, CSF, MI, CMI) discussed in the present work and
to probe the Meissner currents and vortex structure.
In addition, the tricritical point found in our study, and the nature of
the CSF-MI transition could be put to the test in such experiments.

\begin{acknowledgments}
 We thank Thierry Giamarchi, Masaaki Nakamura, Masaki Oshikawa and
 Alexandru Petrescu for fruitful discussions.
 We acknowledge the support of the DARPA-OLE program, of the Swiss
 National Science Foundation under MaNEP and Division II, and of
 a grant from the European Research Council (ERC-319286 QMAC).
\end{acknowledgments}

\appendix
 \section{Mean-field analysis to the long-wave-length effective
 Hamiltonian} 
 \label{sec:MF}
 Here we present the mean-field approach to determine the chemical
 potential in the effective Hamiltonian given by the long-wave-length
 approximation. 

 \subsection{Large magnetic flux case}
 First let us see the case of the large magnetic flux, in which the
 single-particle spectrum forms the double minima in the lower energy
 band.
 Based on the Hamiltonian~(\ref{eq:kinetic--large_flux})
 and~(\ref{eq:interaction--large_flux}) derived by the long-wave-length
 approximation the mean-field energy per site, in which the quantum
 fluctuations are ignored, is assumed to be
 \begin{align}
  E_{\mathrm{MF}}
  &
  =-\frac{1}{2}\left(\mu+E_{0}+\frac{U}{2}-UV_{+}^2V_{-}^2\right)
    \left(\tilde{n}_{+}+\tilde{n}_{-}\right)
  \nonumber \\
  & \quad
    +\frac{U\left(1+2V_{+}^2V_{-}^{2}\right)}{8}
     \left(
      \tilde{n}_{+}
      +\tilde{n}_{-}
     \right)^2
  \nonumber \\
  & \quad 
    +\frac{U\left(1-6V_{+}^2V_{-}^2\right)}{8}
     \left(
      \tilde{n}_{+}
      -\tilde{n}_{-}
     \right)^2,
 \end{align}
 where $\tilde{n}_{\pm}=\langle{\tilde{n}_{\pm,j}}\rangle$.
 From the mean-field energy, the mean-field equations for the density of
 the bosons populating at each band minima are derived by 
 $\partial{E_{\mathrm{MF}}}/{\partial{\tilde{n}_{\pm}}}=0$,
 and lead to the mean-field solution
 $\nbar=\tilde{n}_{+}=\tilde{n}_{-}$ with
 \begin{align}
  & \mu
  = -E_{0}
    +U\left(\nbar-\frac{1}{2}\right)
    +V^{2}_{+}V^{2}_{-}U\left(2\bar{n}+1\right),
  \label{eq:MF-equation1}
 \end{align}
 which determines the chemical potential for the given density
 $\tilde{n}_{\pm}$.
 The obtained mean-field density $\nbar$ can be associated with those of
 the chains, $\nbar_{p}=\langle{n_{j,p}}\rangle$ ($p=1,2$), in the
 original representation.
 The approximated form of the densities~(\ref{eq:density}) leads to
 \begin{align}
  \nbar_{1}=V^{2}_{-}\tilde{n}_{+}+V^{2}_{+}\tilde{n}_{-},
  \nonumber \\
  \nbar_{2}=V^{2}_{+}\tilde{n}_{+}+V^{2}_{-}\tilde{n}_{+},
 \end{align}
 and $\nbar_{1}=\nbar_{2}=\bar{n}$ is immediately concluded since
 $\tilde{n}_{\pm}=\bar{n}$ and $V^{2}_{+}+V^{2}_{-}=1$.

 In addition, the stability of the mean-field solution is confirmed by
 the positive definiteness of Hessian matrix
 $H_{\alpha,\beta}=\partial^{2}{E_{\mathrm{MF}}}/\partial{\tilde{n}_{\alpha}}\partial{\tilde{n}_{\beta}}>0$
 ($\alpha,\beta=\pm$).
 From the straightforward calculation of the eigenvalues of the Hessian
 matrix, the condition of the stable mean-field solution is found to be
 reduced to
 \begin{equation}
  \left(\frac{J_{\perp}}{2J}\right)^2
   <\frac{2\sin^4\left(\phi/2\right)}{3-2\sin^2\left(\phi/2\right)}.
 \end{equation}
 This condition needs the smaller rung hopping as $\phi$ decreases. 
 For example, for the largest magnetic flux case $\phi=\pi$, it leads to
 $J_{\perp}^{2}<8J^2$, and for the less flux $\phi=\pi/2$,
 $J_{\perp}^2<J^2$ is needed.

 \subsection{Small magnetic flux case}

 Next we consider the small magnetic flux case, in which the low-energy 
 single-particle spectrum shows a single minimum in the bottom of the
 lower energy band.
 Neglecting the quantum fluctuations in the approximated long-wave-length
 Hamiltonian~(\ref{eq:Hamiltonian2}), the mean-field energy is given by
 \begin{align}
  E_{\mathrm{MF}}
  =-\frac{1}{2}\left(\mu+E_{0}+\frac{U}{2}\right)\tilde{n}
   +\frac{U}{8}\tilde{n}^2,
 \end{align}
 where $\tilde{n}=\langle{\tilde{n}_{j}}\rangle$ in
 Eq.~(\ref{eq:Hamiltonian2}). 
 Thus the mean-field solution is given by
 $\partial{E_{\mathrm{MF}}}/\partial{\tilde{n}}=0$, which is 
 \begin{equation}
  \mu
   = -E_0+\frac{U}{2}\left(\tilde{n}-1\right).
   \label{eq:MF-equation2}
 \end{equation}
 In addition, from the second-order derivative of the mean-field energy,
 the above mean-field solution is immediately found to be stable. 
 The mean density on the chains is balanced as in
 Eq.~(\ref{eq:density_operator2}), i.e.,
 $\langle{n_{j,1}}\rangle=\langle{n_{j,2}}\rangle=\tilde{n}/2$. 
 Thus, supposing the balanced mean density on the chains to be $\nbar$,
 this density is controlled by the chemical potential as 
 \begin{align}
  \nbar
  =\frac{1}{U}\left(\mu+E_{0}+\frac{U}{2}\right).
  \label{eq:MF-equation3}
 \end{align}

\bibliographystyle{apsrev4-1}

%

\end{document}